\documentclass[review,a4paper,12pt]{elsarticle}
\usepackage{amssymb}
\usepackage{lineno,hyperref}
\usepackage{epsfig}
\usepackage{epstopdf}
\usepackage{graphicx}

\modulolinenumbers[5]

\journal{Journal of Nucl. Instrum. Meth. B}

\begin{document}

\title{
New charge exchange model of GEANT4 \\
for $^{9}$Be(p,n)$^{9}$B reaction
}

\author{Jae Won Shin}
% \ead{shine8199@skku.edu}
\address{Department of Physics, Sungkyunkwan University,
Suwon 440-746, Korea}
\author{Tae-Sun Park}
\ead{tspark@kias.re.kr}
\address{Department of Physics, Sungkyunkwan University,
Suwon 440-746, Korea}

\date{October 5, 2014}

\begin{abstract}
A new data-based charge exchange model of GEANT4 
dedicated to the $^{9}$Be(p,n)$^{9}$B reaction is developed
by taking the ENDF/B-VII.1 differential cross-section data as input. 
Our model yields results that are in good agreement 
with the experimental neutron yield spectrum data obtained
for proton beams of energy $(20\sim35)$ MeV. 
In particular, 
in contrast to all the considered GEANT4 hadronic models,
the peak structure 
resulting from the discrete neutrons
generated by the charge-exchange reaction is 
observed to be accurately reproduced in our model.
\end{abstract}


\maketitle

\section{Introduction}

Neutron sources play essential roles in various industrial 
and scientific fields.
In obtaining neutron beams with the desired energy spectrum,
one approach that is commonly adopted 
is the use of neutron-emitting radio isotopes, 
such as $^{252}$Cf and Be-coupled $^{241}$Am.
In this method, however, the energies of the neutrons 
are on the order of a few MeV and are determined by the isotope. 
That is, the neutron spectra of $^{252}$Cf and $^{241}$Am-Be 
are smooth curves with average energies of $\sim$ 2.2 MeV and 
$\sim$ 4.5 MeV, respectively, and thus, they are not suitable 
when more energetic neutrons are required. 

A more flexible method is to 
bombard proton beams on a target; in this case, 
neutrons are generated mainly 
through the (p,n) charge exchange reaction.
By adjusting the beam energy, 
the target material and the thickness of the target, 
the resulting neutron spectrum can be controlled to a certain degree.
Beryllium is widely used as the target material
due to its high melting point, good thermal conductivity 
and many other desired features. 
Creating particle transport codes
to accurately reproduce
the neutron energy spectrum of the $^{9}$Be(p,n)$^{9}$B reaction
is thus of great importance.

For proton beams of energy ($20\sim 35$) MeV 
impinged on a 0.1 cm thick beryllium target,
we first performed a comparative study
with
GEANT4 \cite{g4n1, g4n2} 
%FLUKA \cite{fluka1, fluka2}, 
and PHITS \cite{phits0},
and observed that
all the platforms substantially underestimate
the neutron yields (${\cal{Y}}_{n}$).\footnote{\protect
There are also MCNPX~\cite{mcnpx1} simulation results for 
11 MeV protons impinged on a 0.2 cm thick beryllium target~\cite{mcnpEp11}.
The simulations show that
although rather good agreement is achieved overall,
the model overestimates the neutron yield
near the end point
(e.g., $\mbox{E}_n \simeq 9$ MeV at forward angle).
}
This finding may not be too surprising because 
the hadronic models of the platforms have been developed for wide use,
but none of the models is specialized for the $^{9}$Be(p,n)$^{9}$B reaction.

In this work, 
we developed a charge exchange model of GEANT4 dedicated 
to the $^{9}$Be(p,n)$^{9}$B reaction, 
taking the ENDF/B-VII.1 differential cross-section data \cite{endf} as input.
When combined with the G4BinaryCascade \cite{BICref1} model for continuum neutrons, 
the developed model is observed to accurately describe
the experimental neutron yield spectra, see Figs. \ref{fig8} and \ref{fig10}.
In particular,
the peak structure
due to the discrete neutrons 
is well reproduced,
whereas all other considered models are highly problematic in this respect.

\section{Simulation tool}
%\section{Results}

GEANT4 is a tool kit 
that allows for microscopic Monte Carlo simulations of 
particles interacting with materials.
The platform has been thoroughly tested and is widely used in
many different scientific fields, such as 
medical physics \cite{G4Med2, Shin1, G4Med3},
accelerator-based radiation studies \cite{ShinAcc1, Shin2, G4Acc2}, 
neutron shielding studies \cite{G4shield2, Bak, G4shield3}, and
environment radiation detection \cite{G4Det2, G4Det3, Shin4}.

In this work, we simulated the neutrons produced by proton beams 
on a $^{9}$Be target by using GEANT4 v10.0.
For the electromagnetic processes, 
we adopted 
``G4EmStandardPhysics$\_$option3".
For the hadronic inelastic processes, 
four different hadronic models were considered: 
``G4BertiniCascade" \cite{BERTIref3}, 
``G4BinaryCascade" \cite{BICref1}, 
``G4Precompound" \cite{PRECOMref1} and 
``G4INCLCascade" \cite{inclPref},
which will be hereafter referred to as
``G4BERTI", ``G4BC", ``G4PRECOM" and ``G4INCL", respectively.
The models are described in detail in the Physics Reference Manual \cite{g4_physRef} 
and Refs. \cite{G4HadPhy1, G4HadPhy2}.

\section{Results}

\subsection{Benchmarking simulations}

\begin{figure}[tbp]
\begin{center}
\epsfig{file=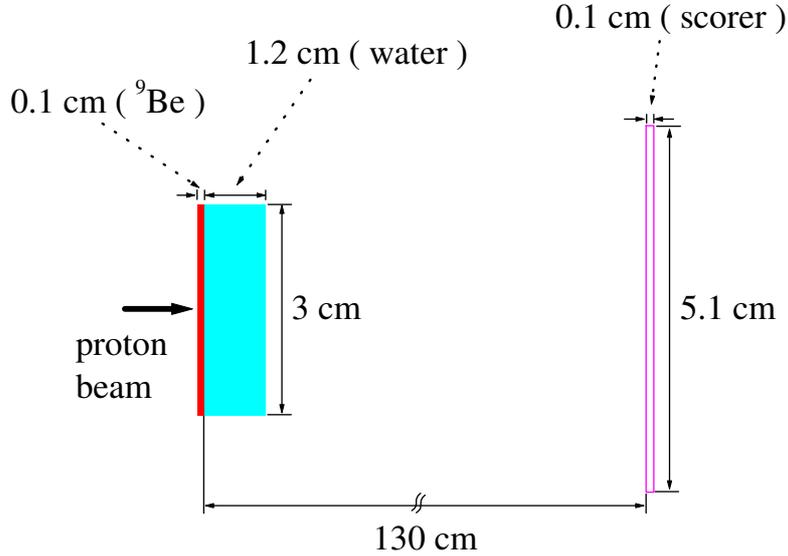, width=4.5in}
\end{center}
\caption{(Color online) Schematic diagram of the simulation geometry.} 
\label{fig1}
\end{figure}

As mentioned in the Introduction, GEANT4 is not equipped with 
a specialized routine for the $^{9}$Be(p,n)$^{9}$B reaction.
To quantify the accuracy of the hadronic models of GEANT4, 
we simulated the neutron yields 
due to 35 MeV protons directed toward a 0.1 cm thick Be target, 
whose experimental data are presented in Ref. \cite{bench4} 
and in the EXFOR database \cite{exfor}. 
According to the experimental setup,
the diameter of the proton beam was set to 0.4 cm (with a  flat shape),
and we placed a cylindrical scoring geometry measuring 0.1 cm in thickness 
and 5.1 cm in diameter
at a distance of 1.3 m from the target; 
see Fig. \ref{fig1} for the simulation geometry.
We also took into account the proton stopper of the 
experiment, {\it i.e.}, the 1.2 cm thick water layer placed behind the target.
We repeated our simulations for each of the four 
hadronic models mentioned 
above.

%%%%%%%%%%%%%%%%%%%%%%%%%%%%%%%%%%%%%%%%%%%%%%%%%%%%%
%%%%%%%%%%%%%%%%%%%%%%%%%%%%%%%%%%%%%%%%%%%%%%%%%%%%%%%

%%%%%%%%%%%%%%%%%%%%%%%%%%%%%%%%%%%%%%%%%%%%%%%%%%%%%
\begin{figure}[tbp]
\begin{center}
\epsfig{file=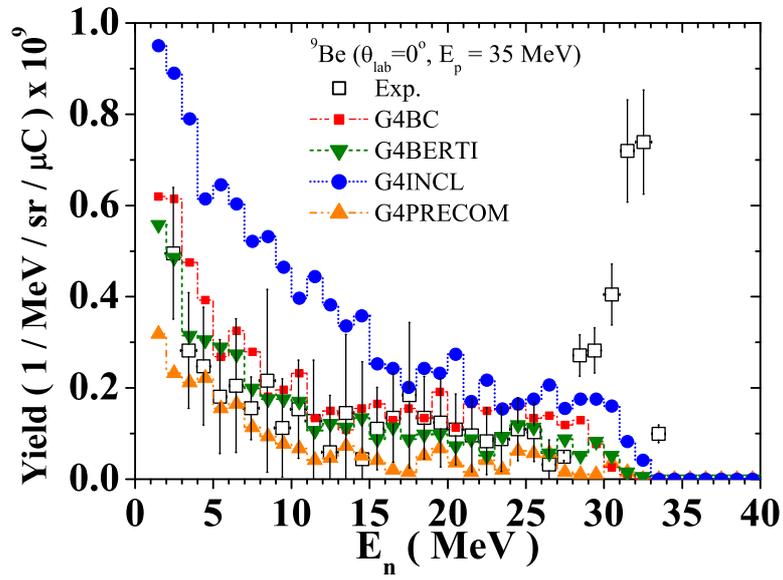, width=4.5in}
\end{center}
\caption{(Color online) Energy spectra of neutrons 
at ${\rm{\theta}}_{\rm{lab}}$ = $0^\circ$ 
produced by 35 MeV protons on a 0.1 cm $^{9}$Be target.
The open squares represent the experimental data \cite{bench4}, 
the red squares G4BC, 
the green inverted triangles G4BERTI, 
the blue circles G4INCL 
and the orange triangles G4PRECOM.} 
\label{fig2}
\end{figure}
%%%%%%%%%%%%%%%%%%%%%%%%%%%%%%%%%%%%%%%%%%%%%%%%%%%%%%%

Figure \ref{fig2} shows the neutron energy spectra 
in the forward angle, $\theta_{\rm lab}=0^\circ$, 
where E$_{\rm{p}}$ and E$_{\rm{n}}$ represent 
for the energy of the incident protons 
and the outgoing neutrons, respectively.
{}The figure clearly shows that the considered models 
do not reproduce the peak structure 
at E$_{\rm{n}}$ $\simeq$ 32 MeV, 
where the peak is mainly due to discrete neutrons produced 
by the $^{9}$Be(p,n)$^{9}$B reaction. 

\begin{figure}[tbp]
\begin{center}
\epsfig{file=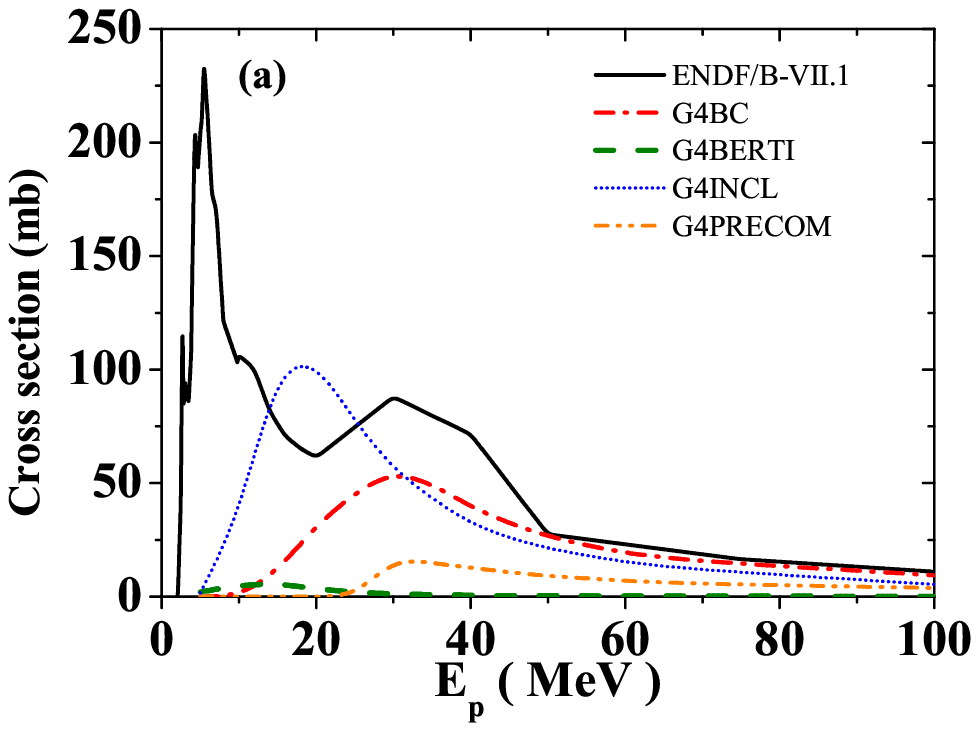, width=4.5in}
\epsfig{file=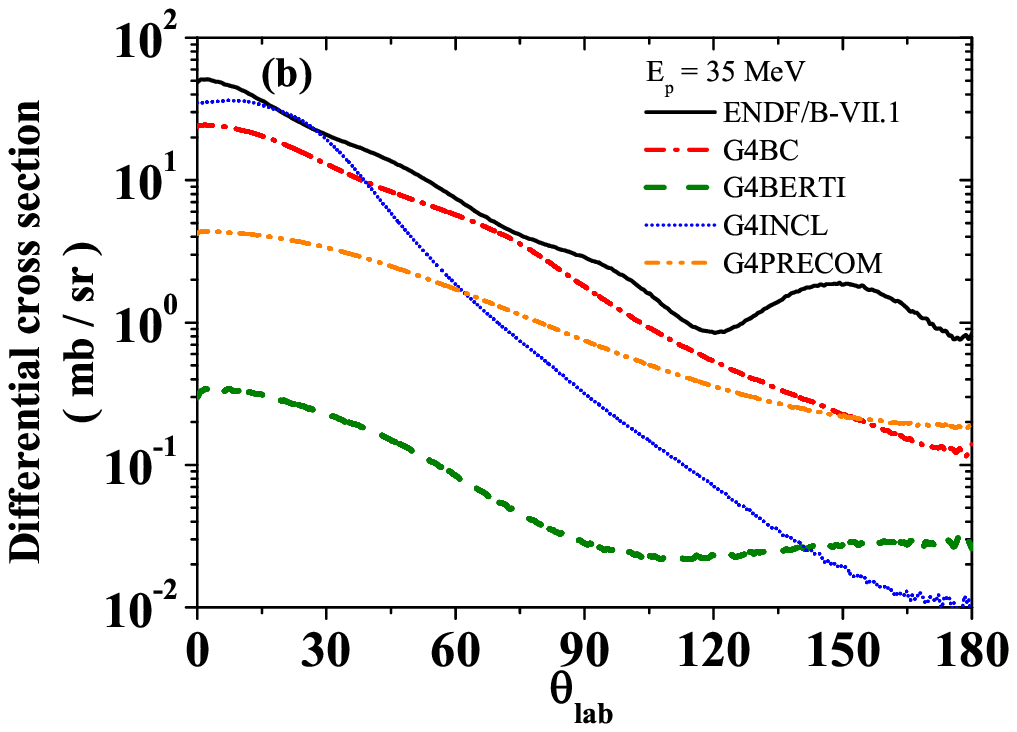, width=4.5in}
\end{center}
\caption{(Color online) Total cross-sections with respect to the incident proton energy (a) 
and the differential cross-sections with respect to the angle ${\rm{\theta}}_{\rm{lab}}$ 
at E$\rm{_{p}}$ = 35 MeV (b) of the $^{9}$Be(p,n)$^{9}$B reaction.} 
\label{fig3}
\end{figure}

\begin{figure}[tbp]
\begin{center}
\epsfig{file=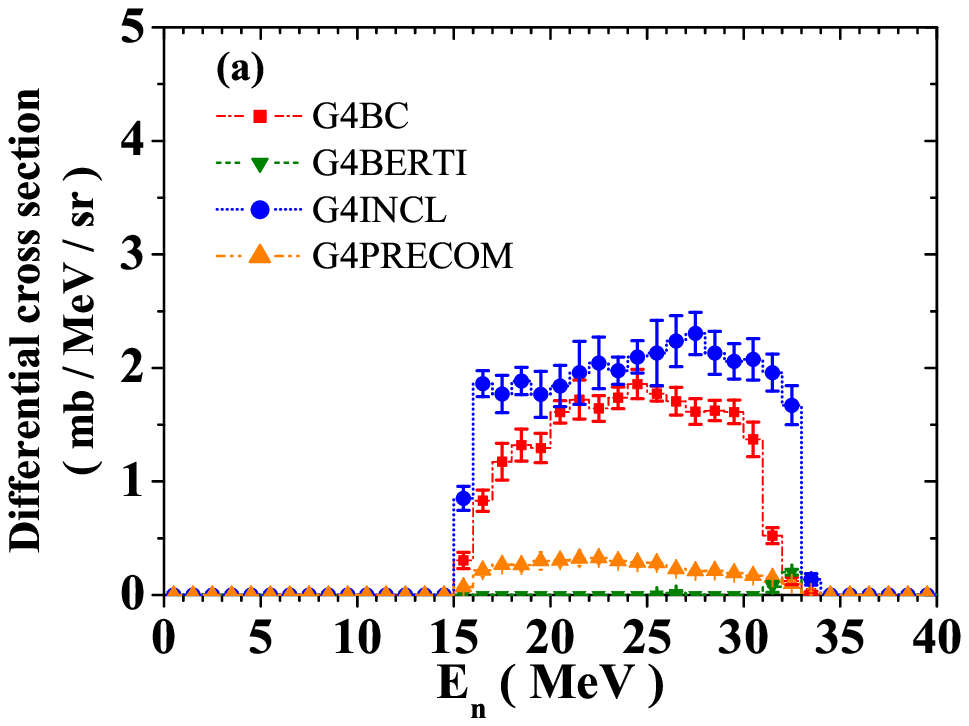, width=4.5in}
\epsfig{file=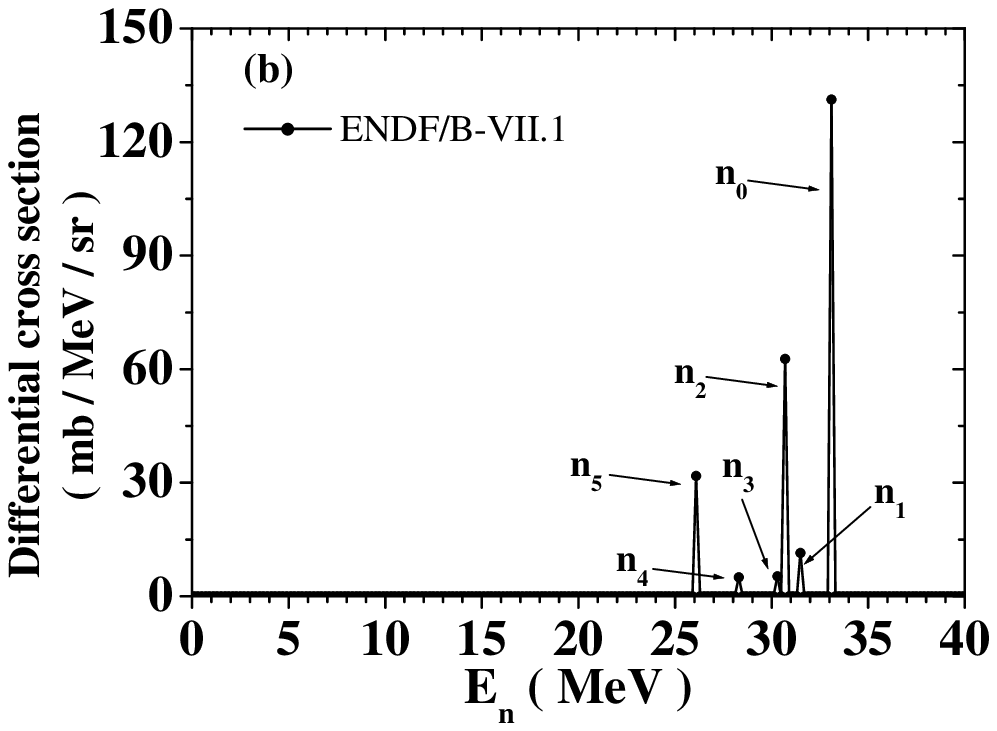, width=4.5in}
\end{center}
\caption{(Color online)
Double differential cross-sections 
with respect to neutron energy 
at $\rm{{\theta}_{lab}}$ = $0^\circ$.
(a) and (b) show the results obtained from the models of GEANT4 and ENDF/B-VII.1, respectively.} 
\label{fig4}
\end{figure}

To confirm our findings, 
we simulated the total and angular differential cross-sections 
of the $^{9}$Be(p,n)$^{9}$B reaction.
The results are presented in Fig. \ref{fig3},
which indicate that
enormous discrepancies exist among the models; indeed, 
none of the models is in agreement with the ENDF/B-VII.1 data.\footnote{
\protect
In extracting the ENDF/B-VII.1 differential cross-section values,
we combined the ENDF MF=3 and MF=6 data using
a software program that is currently under development~\cite{TNudyOld, TNudy}.}
To gain a better microscopic understanding, 
we also plotted
the double differential cross-sections with
respect to the neutron energy in the forward angle, as shown in Fig. \ref{fig4}.
The figure
clearly shows that the GEANT4 hadronic models (Fig. \ref{fig4} (a)) 
fail to 
describe the sharp peak structure of the ENDF/B-VII.1 data (Fig. \ref{fig4} (b)).

%%%%%%%%%%%%%%%%%%%%%%%%%%%%%%%%%%%%%%%%%%%%%%%%%
%%%%%%%%%%%%%%%%%%%%%%%%%%%%%%%%%%%%%%%%%%%%%%%%%%%

Indeed, without the peak structure, 
it is not possible to describe the neutron spectrum accurately.
Therefore, 
we constructed a GEANT4 hadronic model dedicated to the $^{9}$Be(p,n)$^{9}$B reaction,
a detailed description of which is presented 
in the next section.

%%%%%%%%%%%%%%%%%%%%%%%%%%%%%%%%%%%%%%%%%%%%%%%%%
%%%%%%%%%%%%%%%%%%%%%%%%%%%%%%%%%%%%%%%%%%%%%%%%%%%

\subsection{$^{9}$Be(p,n)$^{9}$B charge exchange model}

%%%%%%%%%%%%%%%%%%%%%%%%%%%%%%%%%%%%%%%%%%%%%%%%%%%
\begin{figure}[tbp]
\begin{center}
\epsfig{file=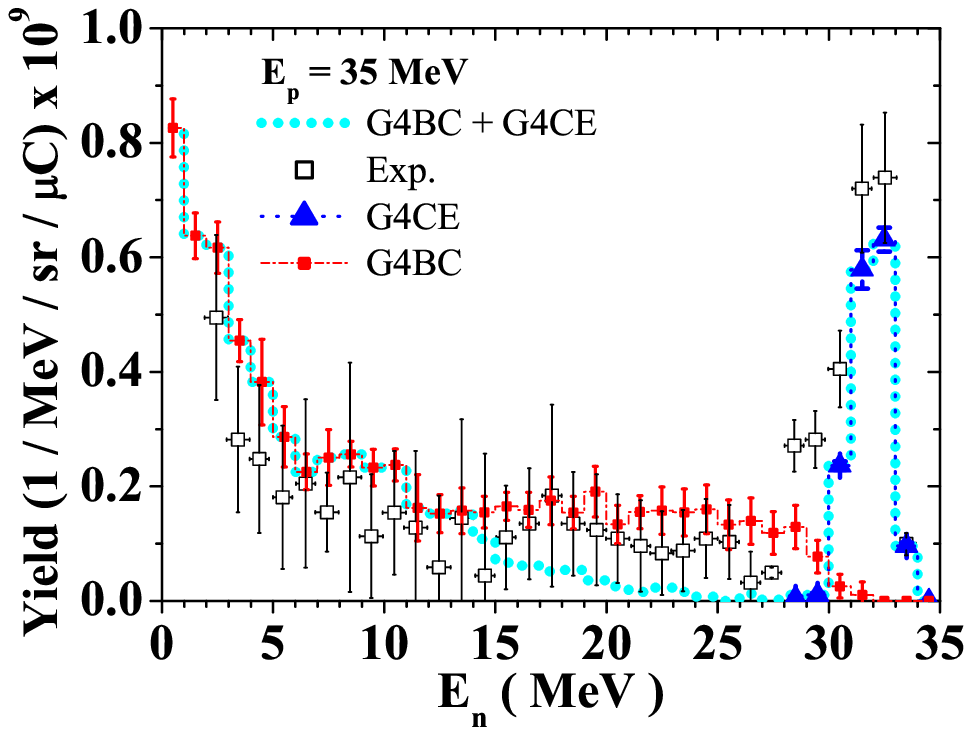, width=4.5in}
\end{center}
\caption{(Color online) 
Energy spectra of neutron yields at ${\rm{\theta}}_{\rm{lab}}$ = $0^\circ$ for 
35 MeV protons on a 0.1 cm $^{9}$Be target.
The open squares, the blue triangles and the red squares 
represent the experimental data \cite{bench4}, the G4CE results and the G4BC results, respectively. 
The dotted line represents the values of ${\cal{Y}}_{n}$ calculated using G4BC + G4CE.
}
\label{fig5}
\end{figure}
%%%%%%%%%%%%%%%%%%%%%%%%%%%%%%%%%%%%%%%%%%%%%%%%%

It should be noted that there is a GEANT4 hadronic model for 
charge exchange reactions, G4ChargeExchange (``G4CE").
To demonstrate the accuracy of the model, 
we repeated the simulation under the same 
conditions described above.
Noting that G4CE covers only the charge-exchange reaction
but not other continuum neutrons,
we studied three cases, G4CE, G4BC and G4BC+G4CE.
For the G4BC+G4CE case, 
to avoid any possible double counting,
we removed
the neutrons produced by the $^{9}$Be(p,n)$^{9}$B reaction of G4BC
for the entire energy region
by making use of the G4UserSteppingAction class.
The resulting neutron yields are plotted in Fig. \ref{fig5},
which shows that
adding G4CE on top of G4BC
improves the accuracy remarkably.
However, an error of approximately 20\% in the height 
for the n$_{0}$ peak remains, and the n$\rm{_i}$ (i $\geq$ 1) peaks are still missing,
where n$_{0}$ and n$\rm{_i}$ denote the neutrons with the residual 
$^{9}$B in the ground and i-th excited states, respectively.

%%%%%%%%%%%%%%%%%%%%%%%%%%%%%%%%%%%%%%%%%%%%%%%
\begin{figure}[tbp]
\begin{center}
\epsfig{file=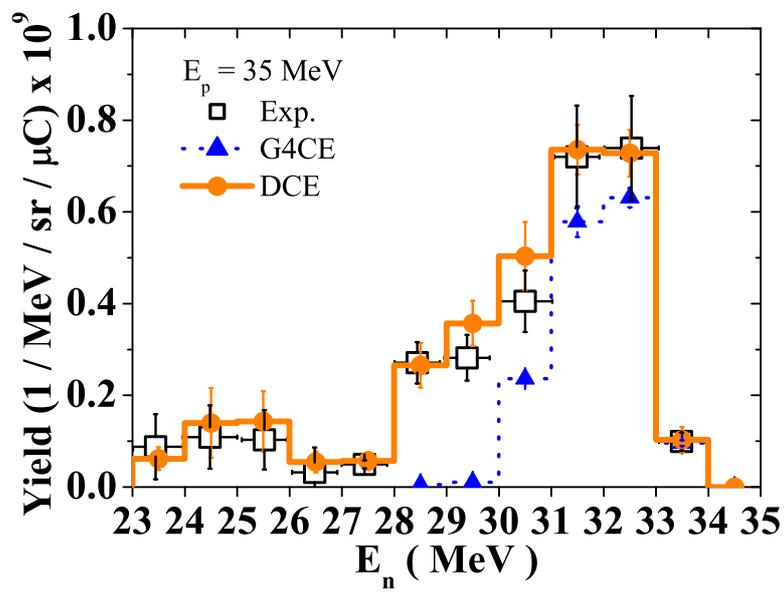, width=4.5in}
\end{center}
\caption{(Color online) Neutron energy spectra at ${\rm{\theta}}_{\rm{lab}}$ = $0^\circ$ 
produced by 35 MeV protons.
The open squares represent the experimental data \cite{bench4}, 
the blue triangles with a dotted line represent the G4CE results, 
and the orange circles with a solid line represent the results of our DCE model.}
\label{fig6}
\end{figure}
%%%%%%%%%%%%%%%%%%%%%%%%%%%%%%%%%%%%%%%%%%%%%%%%%

For an accurate description of discrete neutrons, 
we developed a data-based charge exchange (DCE) model.
For the discrete neutrons from the $^{9}$Be(p,n)$^{9}$B reaction, 
the ENDF/B-VII.1 differential cross-section data \cite{endf} 
of the reaction were taken as input.
The resulting ${\cal{Y}}_{n}$ is plotted in Fig. \ref{fig6}, 
which shows that the prediction of the DCE model is in good agreement 
with the experimental data.

%%%%%%%%%%%%%%%%%%%%%%%%%%%%%%%%%%%%%%%%%%%%%%%%%%%%%%%%%%%
\begin{figure}[tbp]
\begin{center}
\epsfig{file=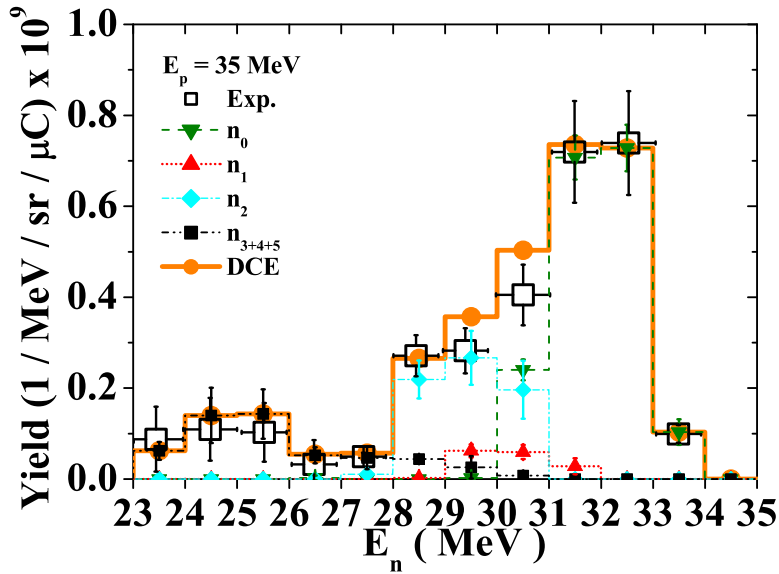, width=4.5in}
\end{center}
\caption{(Color online) Neutron energy spectra 
at ${\rm{\theta}}_{\rm{lab}}$ = $0^\circ$ 
produced by 35 MeV protons on a $^{9}$Be target.
The open squares represent the experimental data \cite{bench4}.
The green inverted triangles, the red triangles, the cyan diamonds and the black squares 
denote the calculated values of n$_{0}$, n$_{1}$, n$_{2}$
and n$_{3}+$n$_{4}+$n$_{5}$,
respectively,
and the orange circles with a solid line represent the results of our DCE model.}
\label{fig7}
\end{figure}
%%%%%%%%%%%%%%%%%%%%%%%%%%%%%%%%%%%%%%%%%%%%%%%%%

The contribution of each n$\rm{_i}$ to the yield is plotted in Fig. \ref{fig7}, 
which shows that the contribution of n$_{2}$ is quite substantial, without 
which the shoulder of the peak cannot be reproduced.
In contrast, the contributions of 
n$_{1}$ and n$\rm{_i}$, with i $\geq$ 3, are observed to be marginal.
The figure also shows that the width of each peak is approximately 3 MeV,  
which is mainly due to the energy loss of the incident protons. 
For 35 MeV protons, 
the calculated average energy loss of the incident protons 
in the 0.1 cm thick $^{9}$Be target is 2.58 $\pm$ 0.13 MeV, 
and the attenuation of the neutron energy 
due to the 1.2 cm thick water placed behind the target is approximately 0.2 MeV. 

%%%%%%%%%%%%%%%%%%%%%%%%%%%%%%%%%%%%%%%%%%%%%
\begin{figure}[tbp]
%\begin{center}
\epsfig{file=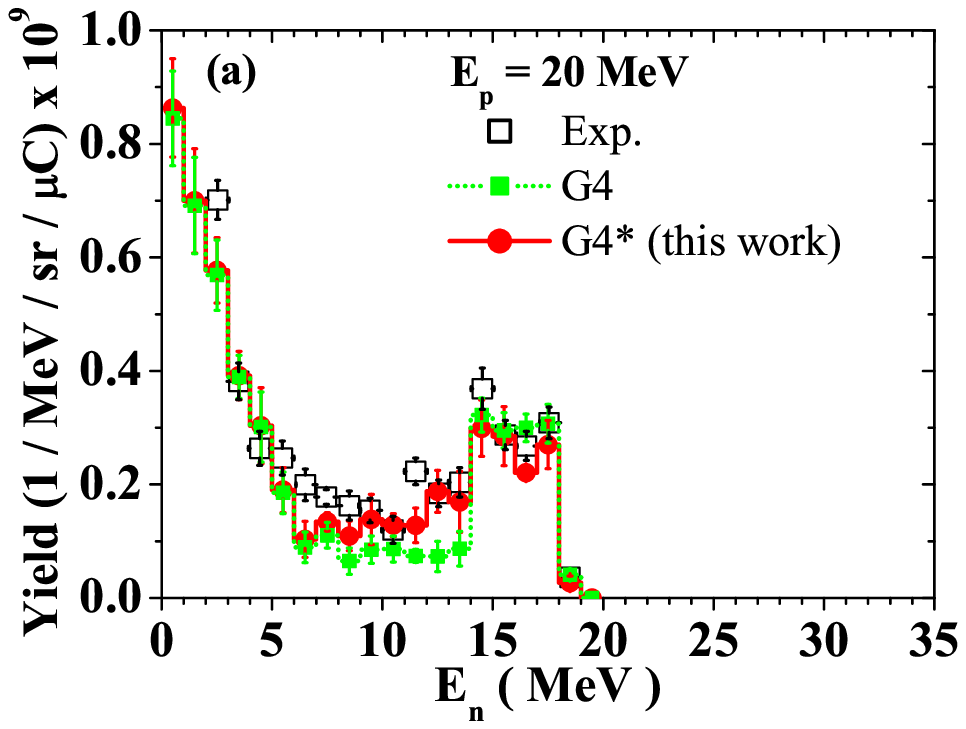, width=2.7in}
\epsfig{file=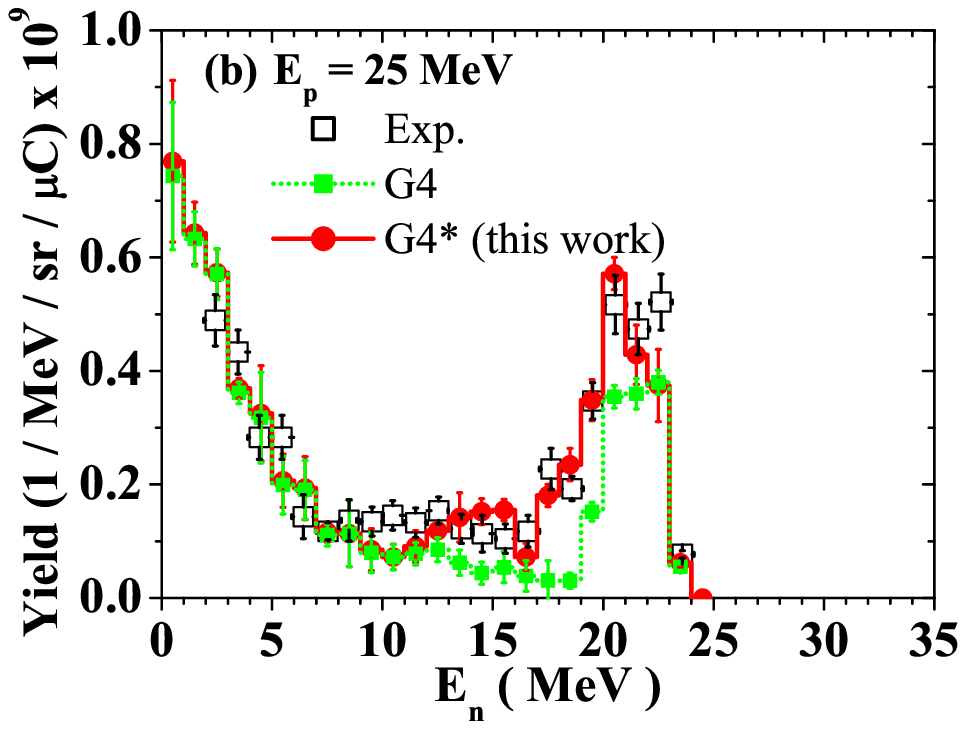, width=2.7in}
\epsfig{file=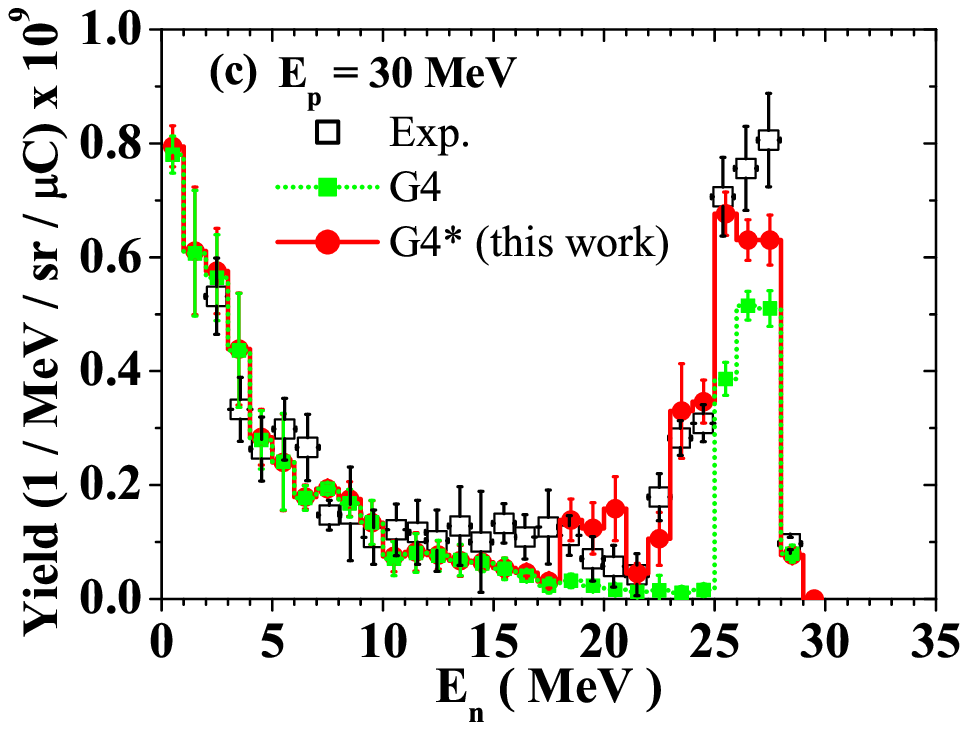, width=2.7in}
\epsfig{file=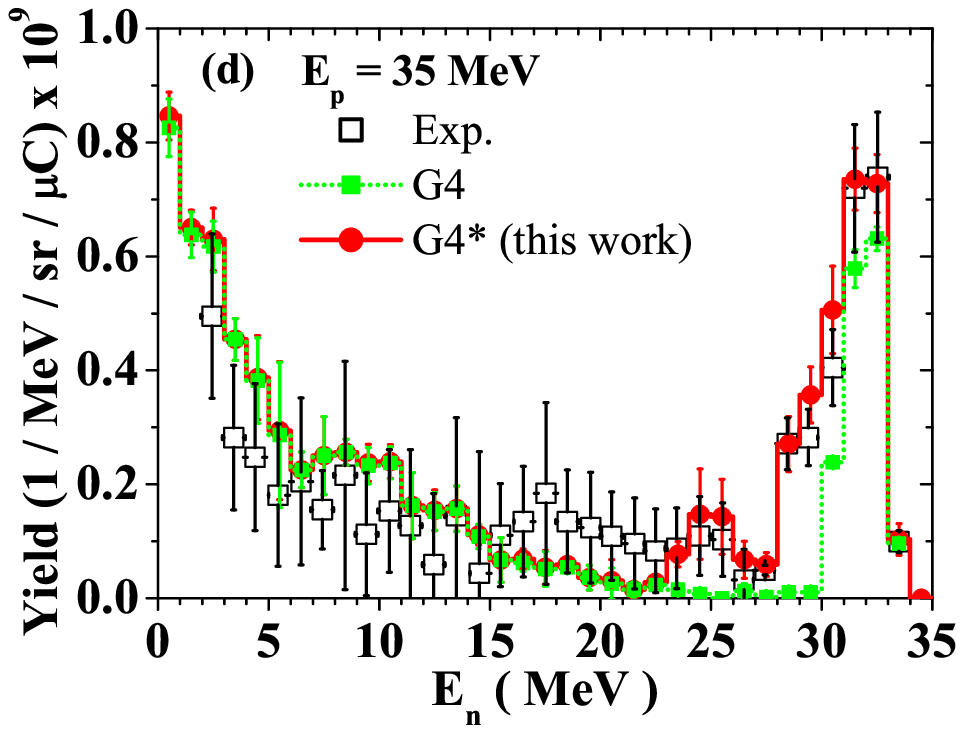, width=2.7in}
%\end{center}
\caption{(Color online) 
Neutron energy spectra at ${\rm{\theta}}_{\rm{lab}}$ = $0^\circ$ 
produced by 20, 25, 30 and 35 MeV incident proton beams on a $^{9}$Be target
are plotted in (a), (b), (c) and (d), respectively.
The open squares shown in black are the experimental data \cite{bench4}.
The green squares with dotted lines and the red circles with solid lines denote
the values of ${\cal{Y}}_{n}$ calculated using G4 (G4BC + G4CE) and G4* (G4BC + DCE), respectively.} 
\label{fig8}
\end{figure}
%%%%%%%%%%%%%%%%%%%%%%%%%%%%%%%%%%%%%%%%%%%%%%%%

%%%%%%%%%%%%%%%%%%%%%%%%%%%%%%%%%%%%%%%%%%%%%%%
\begin{table}
\caption{The calculated-to-experimental ratio
%\cite{bench4} 
for the total neutron yields. 
%G4 and G4* represent the neutron yields calculated using G4BC + G4CE and G4BC + DCE, respectively.
}
%\begin{ruledtabular}
\begin{tabular}{c|cccc}
%\colrule
%  E$\rm{_{p}}$ (MeV) & G4BC/Exp.      & G4/Exp.         & G4*/Exp.        &\\ \hline
  E$\rm{_{p}}$ (MeV) & G4BC      & G4BC+G4CE         & this work        &\\ \hline
        20      & 0.67 $\pm$ 0.05 & 0.77 $\pm$ 0.03 & 0.84 $\pm$ 0.04 &\\
        25      & 0.70 $\pm$ 0.05 & 0.71 $\pm$ 0.03 & 0.99 $\pm$ 0.04 &\\
        30      & 0.67 $\pm$ 0.04 & 0.67 $\pm$ 0.03 & 0.97 $\pm$ 0.04 &\\
        35      & 0.82 $\pm$ 0.04 & 0.84 $\pm$ 0.04 & 1.13 $\pm$ 0.05 &\\
\end{tabular}
%\end{ruledtabular}
\label{table_1}
\end{table}
%%%%%%%%%%%%%%%%%%%%%%%%%%%%%%%%%%%%%%%%%%%%%%%

In addition to the discrete neutrons that are responsible for the peaks, 
continuum neutrons also appear in the low-energy region.
Figure \ref{fig2} and Fig. \ref{fig5} show that the latter can be well described by G4BC.
We thus combined G4BC with our DCE model to cover the neutron yield spectrum for 
the entire energy region; the resulting G4BC + DCE model is referred to as G4*.
The results of our combined G4* model for the 
neutron spectra produced by 20, 25, 30 and 35 MeV proton 
beams impinging on a 0.1 cm thick $^{9}$Be target 
at $\theta_{\rm lab}=0^\circ$
are plotted 
in Fig. \ref{fig8}.
A comparison of this figure with Fig. \ref{fig2} reveals that the agreement with the experimental data 
near the peak region 
-- where the $^{9}$Be(p,n)$^{9}$B reaction plays a dominant role -- 
is significantly improved. 
Below the peak region, 
the models show a tendency to 
underestimate the neutron yields, 
although their predictions are within the error of the data; 
see, for example, the E$\rm{_{n}}$ = (15 $\sim$ 24) MeV region 
for E$\rm{_{p}}$ = 35 MeV plotted in Fig. \ref{fig8} (d). 
This discrepancy 
may derive from the inaccuracy in treating the 
continuum neutrons 
emanating from other reaction channels, such as 
$^{9}$Be(p,pn)$^{8}$Be and $^{9}$Be(p,n$\alpha$)$^{5}$Li. 

The calculated-to-experimental (C/E) ratios
for the total 
and peak neutron yields are tabulated 
in Table \ref{table_1} and Table \ref{table_2}, respectively.
Table \ref{table_1} shows that
the error in the C/E ratio of G4BC for the total
neutron yield is approximately 30 \%,
which is not particularly different from the error yielded by the G4CE model.
However, when our DCE model is added, the error is reduced to approximately
$(0\sim 16)\ \%$.
The improvement for the peak neutrons
is observed to be more dramatic.
That is, 
the results of G4BC 
-- which hardly covers the discrete neutrons 
that are responsible for the peaks --
are completely 
unacceptable.
By adding the G4CE model on top of the G4BC model, 
the error in the C/E ratio is reduced to $(21\sim 48)\ \%$.
Moreover, by replacing the G4CE model with our DCE model,
the error is further reduced to $(4\sim 16)\ \%$.

%%%%%%%%%%%%%%%%%%%%%%%%%%%%%%%%%%%%%%%%%%%%%%%
\begin{table}
\caption{The calculated-to-experimental ratio
for the peak neutron yield,
where the peak region of each E$_{\rm p}$ is
denoted in the 2nd column.
}
%\begin{ruledtabular}
\begin{tabular}{c|cc|cccc}
%\colrule
% E$\rm{_{p}}$ (MeV) & E$\rm{_{n}}$ (MeV) & & G4BC/Exp.      & G4/Exp.         & G4*/Exp.        &\\ 
  E$\rm{_{p}}$ (MeV) & E$_{\rm n}$ (MeV) & &  G4BC      & G4BC+G4CE         & this work        &\\
               & (peak region) & &                &                 &                 &\\ \hline
            20 &  11 $\sim$ 19 & & 0.38 $\pm$ 0.04 & 0.79 $\pm$ 0.04 & 0.84 $\pm$ 0.06 &\\
            25 &  17 $\sim$ 24 & & 0.20 $\pm$ 0.03 & 0.59 $\pm$ 0.02 & 0.96 $\pm$ 0.05 &\\
            30 &  22 $\sim$ 29 & & 0.14 $\pm$ 0.03 & 0.52 $\pm$ 0.02 & 0.94 $\pm$ 0.04 &\\
            35 &  28 $\sim$ 34 & & 0.10 $\pm$ 0.03 & 0.67 $\pm$ 0.02 & 1.16 $\pm$ 0.06 &\\
\end{tabular}
%\end{ruledtabular}
\label{table_2}
\end{table}
%%%%%%%%%%%%%%%%%%%%%%%%%%%%%%%%%%%%%%%%%%%%%%%

%%%%%%%%%%%%%%%%%%%%%%%%%%%%%%%%%%%%%%%%%%%%%%%%%%%%%%%
\begin{figure}[tbp]
\begin{center}
\epsfig{file=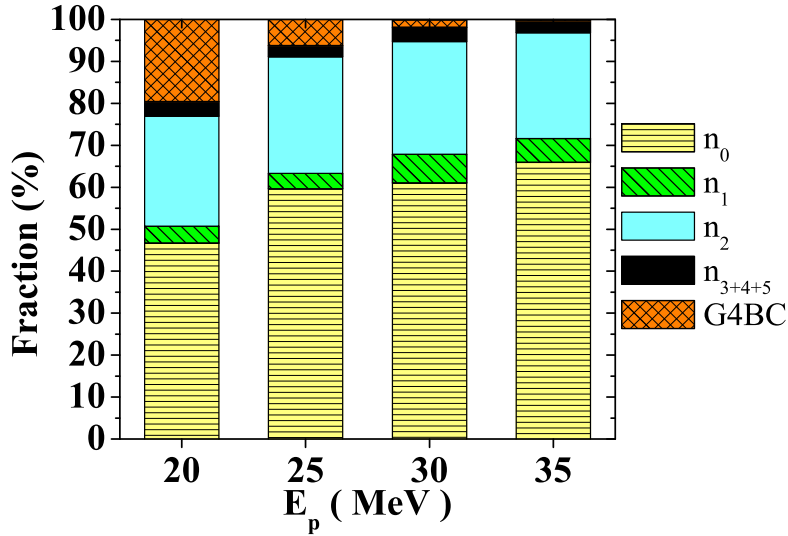, width=4.5in}
\end{center}
\caption{(Color online) Fraction of each contribution for the calculated peak neutron yield.}
\label{fig9}
\end{figure}
%%%%%%%%%%%%%%%%%%%%%%%%%%%%%%%%%%%%%%%%%%%%%%%%%%%%%%%%%

The fraction of each peak contribution 
is plotted 
in Fig. \ref{fig9}, 
which shows the relative importance of the peaks.
At E$\rm{_{p}}$ = 35 MeV, 
the fractions of n$_{0}$, n$_{1}$, n$_{2}$ and n$_{3+4+5}$ 
are approximately 66\%, 5.6\%, 25\% and 3\%, respectively, 
and the G4BC contribution is observed to be negligibly small ($\sim$ 0.3\%).
The importance of continuum neutrons, however, increases at low 
proton beam energies, covering approximately 20\% of the peak yield at E$\rm{_{p}}$ = 20 MeV.

%%%%%%%%%%%%%%%%%%%%%%%%%%%%%%%%%%%%%%%%%%%%%%%%%%%%%%%
\begin{figure}[tbp]
\begin{center}
\epsfig{file=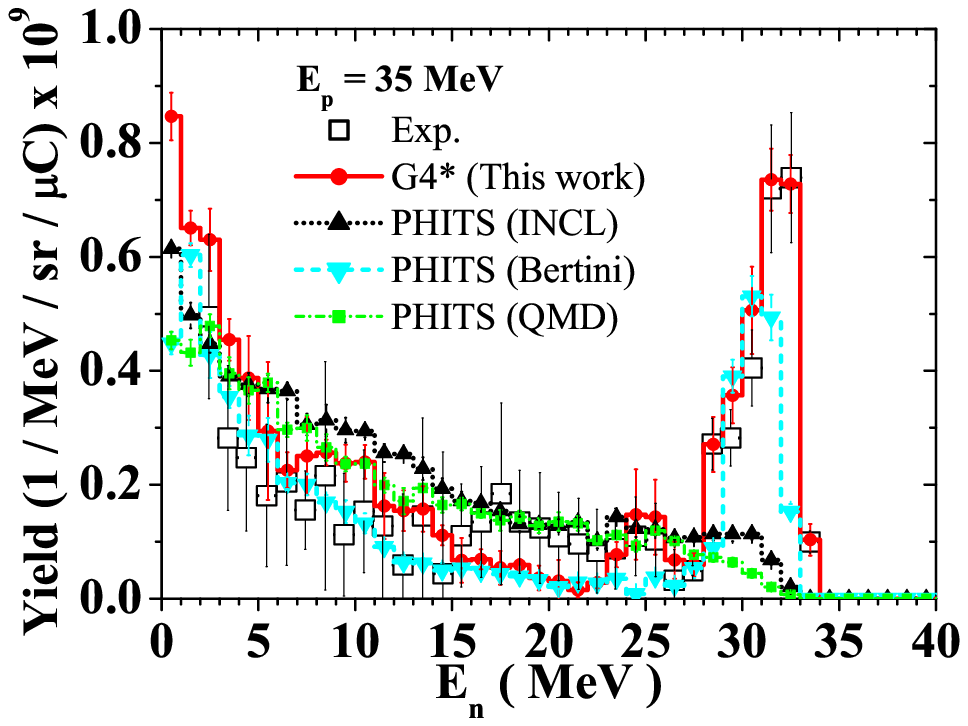, width=4.5in}
\end{center}
\caption{(Color online) 
Neutron energy spectra at ${\rm{\theta}}_{\rm{lab}}$ = $0^\circ$ 
for 35 MeV protons on a $^{9}$Be target.
The open squares shown in black represent the experimental data \cite{bench4}, and
the red solid line with circles represents the G4* (G4BC + DCE) results. 
The black dotted line with triangles, 
the cyan dashed line with inverted triangles and 
the green long-dashed line with squares represent the results obtained by 
the PHITS calculation
using the INCL, Bertini and QMD models, respectively.}
\label{fig10}
\end{figure}
%%%%%%%%%%%%%%%%%%%%%%%%%%%%%%%%%%%%%%%%%%%%%%%%%%%%%%%%%

We also performed 
PHITS simulations under the same conditions, 
exploring three hadronic models of PHITS,
INCL \cite{Pmodels, Pincl}, 
Bertini \cite{Pmodels} 
and QMD \cite{Pmodels}. 
The results are plotted in Fig. \ref{fig10}.
Below the peak region (e.g., E$\rm{_{n}}$ $\lesssim$ 24 MeV), 
all the models are consistent with the data, 
taking into account the large experimental error bars.
However, 
only the Bertini model of PHITS reproduces the peak structure,
and the general behavior of the model
is rather similar to that of the G4BC+G4CE model.

\section{Conclusion}

We examined several hadronic models of GEANT4 
for the neutrons produced by proton beams impinging on a $^{9}$Be target 
and observed that none of the models reproduces the peak structure 
of the neutron spectrum.
Because the peak structure is due to the discrete neutrons 
generated by the $^{9}$Be(p,n)$^{9}$B reaction, 
this finding suggests that the reaction is not properly implemented 
in the models considered.

The charge exchange model of GEANT4, G4CE, 
was also studied; it was observed
that the model
reproduces the n$_{0}$ peak, but its height is reduced to approximately 80\%, 
and n$\rm{_i}$ peaks with i $\geq$ 1 remain missing.
To eliminate this discrepancy, 
we developed a new data-based GEANT4 model dedicated to the 
$^{9}$Be(p,n)$^{9}$B reaction by incorporating 
ENDF/B-VII.1 differential cross-section data of the reaction into G4CE.
For proton beams of energy E$\rm{_{p}}$ = (20 $\sim$ 35) MeV,
the resulting model predictions are in good agreement with the 
experimental data.
We also observed that 
noticeable discrepancies persist below the peak region. 
For an accurate reproduction of the neutron yields for 
the entire energy region, 
it is extremely important to extend our work 
to take into account the ENDF data of all the p + $^{9}$Be channels, 
which is currently in progress.

\section*{Acknowledgments}
This work was supported in part by the Basic
Science Research Program through the Korea Research
Foundation (NRF-2011-0025116, NRF-2012R1A1A2007826, NRF-2013R1A1A2063824).

%\section{Bibliography styles}

%\section*{References}

\bibliography{mybibfile}

%\section{Table}

%\section{Figures}

\end{document}